\documentclass[aps,prb,twocolumn,superscriptaddress,a4paper]{revtex4-1}

\usepackage{graphicx}
\usepackage{enumitem}
\usepackage{booktabs}
\usepackage{amsmath}
\usepackage{placeins}
\usepackage[pdftex,colorlinks=true,linkcolor=blue,citecolor=blue,filecolor=blue,baseurl=empty]{hyperref}

\AtBeginDocument{
\heavyrulewidth=.08em
\lightrulewidth=.05em
\cmidrulewidth=.03em
\belowrulesep=.65ex
\belowbottomsep=0pt
\aboverulesep=.4ex
\abovetopsep=0pt
\cmidrulesep=\doublerulesep
\cmidrulekern=.5em
\defaultaddspace=.5em
}

\newcommand{\avg}[1]{\big< #1 \big>} 
\renewcommand{\v}[1]{\ensuremath{\mathbf{#1}}} 
\newcommand{\ket}[1]{\big| #1 \big>} 
\newcommand{\bra}[1]{\big< #1 \big|} 
\newcommand{\matrixel}[3]{\big< #1 \vphantom{#2#3} \big| #2 \big| #3 \vphantom{#1#2} \big>} 
\newcommand{\lr}[1]{\left(#1\right)}

\newcommand{\rec}[1]{\frac{1}{#1}}
\newcommand{\cC}[1]{c^{\dagger}_{#1}}
\newcommand{\cA}[1]{c_{#1}}

\begin{document}

\title[Topological phases in the Extended Hubbard Model]{Topological phases arising from attractive interaction and pair hopping in the Extended Hubbard Model}

\author{Roman Rausch}
\email[]{rausch.roman.72e@st.kyoto-u.ac.jp}
\affiliation{Department of Physics, Kyoto University, Kyoto 606-8502, Japan}

\author{Matthias Peschke}
\email[]{mpeschke@physnet.uni-hamburg.de}
\affiliation{Department of Physics, University of Hamburg, Jungiusstra{\ss}e 9, D-20355 Hamburg, Germany}
\affiliation{Institute for Theoretical Physics Amsterdam and Delta Institute for Theoretical Physics, University of Amsterdam, Science Park 904, 1098 XH Amsterdam, The Netherlands}

\begin{abstract}
The extended Hubbard model with an attractive density-density interaction, positive pair hopping, or both, is shown to host topological phases, with a doubly degenerate entanglement spectrum and interacting edge spins. This constitutes a novel instance of topological order which emerges from interactions. When the interaction terms combine in a charge-SU(2) symmetric fashion, a novel partially polarized pseudospin phase appears, in which the topological features of the spin degrees of freedom coexist with long-range $\eta$-wave superconductivity. Thus, our system provides an example of an interplay between spontaneous symmetry breaking and symmetry-protected topological order that leads to novel and unexpected properties.
\end{abstract}

\maketitle

\section{Introduction}

Spontaneous symmetry breaking and symmetry-protected topological order (SPTO) constitute two major schemes by which phases of matter can be classified. While the former usually requires interactions, the latter is mainly understood in terms of winding numbers of a noninteracting bandstructure. An interplay between the two can be achieved by adding interactions to a topological bandstructure, which alters the corresponding invariants or the nature of the involved edge states \cite{Fidkowski_2010, Yoshida_2014}. Another intriguing question is whether SPTO itself can arise from interactions, with the possibility of novel properties beyond noninteracting band topology, as a result of the richness of interacting systems \cite{Gonzalez-Cuadra_2019A, Gonzalez-Cuadra_2019B}.

One well-established paradigm of an SPTO system is the Kitaev chain \cite{Kitaev_2001} with Majorana edge modes. While it is not, strictly speaking, an interacting Hamiltonian, but rather a quadratic mean-field one which is diagonalizable by a Bogoliubov transformation, some effort has been put into obtaining the same phase from interacting spinless fermions \cite{Kraus_2013, Lang_2015, Iemini_2015b}.

The $S=1$ Haldane spin chain has proven to be another paradigm for an interacting SPTO system and can serve as a helpful guide. The order is evidenced by a two-fold degenerate eigenvalue spectrum of the reduced density matrix (``entanglement spectrum''), a string order parameter, and entangled $S=1/2$ spins localized at the edges of an open chain \cite{Haldane_1983, Kennedy_1990, Mikeska_2004, Pollmann_2010}. When anisotropy or a transverse field is added to the Hamiltonian, a Haldane phase remains robust in a region of the phase diagram, with the entanglement spectrum still being twofold degenerate, while the string order may vanish \cite{Pollmann_2010}.

A straightforward way to generate a Haldane state in an $S=1/2$ system is by coupling pairs of spins to an effective $S=1$, which can be typically achieved by a ferromagnetic interaction. A frustrated $J_1-J_2$ chain hosts a Haldane phase with dimerized spins \cite{Agrapidis_2019}. An alternating ferromagnetic spin-spin coupling on every second site also leads to a Haldane phase in the Hubbard chain, supplanting the Mott phase entirely \cite{Lange_2015}. A related approach involves explicit dimerization of the Hamiltonian \cite{Koudai_2019, Ghelli_2020}. A dimerized topological bond-order phase was recently reported for bosons \cite{Gonzalez-Cuadra_2019A, Gonzalez-Cuadra_2019B}. In terms of fermionic models, a Haldane phase is found in the anisotropic $t-J$ model \cite{Fazzini_2019} or a 3-leg Hubbard ladder at 2/3 filling \cite{Nourse_2016}. A completely different topological phase was found in a model with an attractive triplet-triplet interaction \cite{Keselman_2015, Verresen_2019}, which is similar to ferromagnetic coupling.

In this work, we report the existence of novel topological phases of the spin degrees of freedom in the 1D Hubbard chain extended by an \textit{attractive} density-density coupling and a pair hopping with an \textit{overall positive} coupling constant, which are not dimerized and exhibit notable differences from the Haldane phase, as will be explained below.

\section{Model}

Our model reads as follows:
\begin{equation}
\begin{split}
H &= -t \sum_{\left<ij\right>\sigma} \lr{c^\dagger_{i\sigma} c_{j\sigma} + H.c.}\\
  &\quad+U \sum_{i} \left[ \lr{n_{i\uparrow}-\rec{2}}\lr{n_{i\downarrow}-\rec{2}} +\rec{4}\right] \\
  &\quad+V_{z}/4 \sum_{\left<ij\right>} \lr{n_i-1}\lr{n_j-1}\\
  &\quad-V_{xy}/2\sum_{\left<ij\right>} \lr{ \cC{i\uparrow} \cC{i\downarrow} \cA{j\downarrow} \cA{j\uparrow} + H.c. },
\label{eq:model}
\end{split}
\end{equation}
where $\cC{i\sigma}$ creates an electron with the spin projection $\sigma = \uparrow, \downarrow$ at site $i$ and $n_{i\sigma}=\cC{i\sigma}\cA{i\sigma}$ is the corresponding density, the total density being $n_i=\sum_{\sigma} n_{i\sigma}$. The chemical potential is kept fixed in the Hamiltonian, so that the ground state is mostly found at half filling $N = \sum_i \avg{n_i} = L$ (with $L$ being the length of a 1D chain) except for some superconducting phases (see below). The physical meanings of the bare energy scales are as follows: $t$ is the hopping amplitude between nearest neighbors (denoted by the angle brackets $\left<ij\right>$), $U$ is the on-site Coulomb interaction, $V_{z}$ the nearest-neighbor Coulomb interaction and $V_{xy}$ the pair-hopping amplitude. Note that all these three terms can be derived from the general interaction term under the assumption of constant matrix elements \cite{Strack_1994}. Thus, the model studied here can also be seen as a piece of a larger phase diagram of the extended 1D Hubbard model with nearest-neighbor interactions. The parameter range discussed in this work is $U>0$, $V_{xy}<0$, $V_z<0$. Interestingly, a similar attractive parameter range was discussed as an effective model for DNA duplexes \cite{Starikov_2003}.

Our definition of $V_{z}$ and $V_{xy}$ is slightly different from the usual convention, but natural in terms of the charge-SU(2) symmetry of the model. Namely, defining the pseudospin operators
\begin{equation}
\begin{split}
T^+_i &= \lr{-1}^i \cA{i\downarrow} \cA{i\uparrow} \\
T^-_i &= \lr{-1}^i \cC{i\uparrow} \cC{i\downarrow} \\
T^x_i &= 1/2\lr{T^+_i+T^-_i} \\
T^y_i &= i/2\lr{T^+_i-T^-_i} \\
T^z_i &= 1/2\lr{n_i-1}
\label{eq:Tdef}
\end{split}
\end{equation}
we notice that while spin operators couple $\uparrow$- and $\downarrow$-states, pseudospin interactions couple empty $\ket{0}$ and doubly occupied (``doublon'') sites $\ket{\uparrow\downarrow}$ with the same SU(2) algebra relations \cite{Essler_2005}. $V_z$ couples only the $z$ components and is analogous to an Ising term, while $V_{xy}$ couples the $x$- and $y$-components and introduces doublon hopping.

Using these operators and introducing the holon density $n^h_i = 2n_{i\uparrow}n_{i\downarrow} -n_i+1$, the model can be compactly rewritten as follows:
\begin{equation}
\begin{split}
H &= -t \sum_{\left<ij\right>\sigma} \lr{c^\dagger_{i\sigma} c_{j\sigma} + H.c.} +U/2 \sum_{i} n^h_i \\
  &\quad+V_{z} \sum_{\left<ij\right>} T^z_iT^z_j + V_{xy}/2\sum_{\left<ij\right>} \lr{ T^+_iT^-_j + H.c. }
\label{eq:modelXXZ}
\end{split}
\end{equation}
At the charge-SU(2) symmetric line $V_{xy}=V_z=V$ we can use the vector notation $\v{T}_i=\lr{T^x_i,T^y_i,T^z_i}$:
\begin{equation}
H = -t \sum_{\left<ij\right>\sigma} \lr{c^\dagger_{i\sigma} c_{j\sigma} + H.c.} + U/2 \sum_i n^h_i + V \sum_{\left<ij\right>} \v{T}_i\cdot\v{T}_j.
\label{eq:H_SU2}
\end{equation}
For $V_{xy}=0$, the model is known as the extended Hubbard model \cite{Nakamura_1999, Vojta_1999, Nakamura_2000, Lin_2000, Sengupta_2002, Tsuchiizu_2002, Jeckelmann_2002, Sandvik_2004, Ejima_2007, Kumar_2009, Senechal_2013, Kantian_2019}, for $V_z=0$ as the Penson-Kolb-Hubbard model \cite{Penson_1986, Kolb_1986, Hui_1993, Bhattacharyya_1995, Japaridze_1997, Robaszkiewicz_1999, Japaridze_2001, Japaridze_2002, Kapcia_2016}. We focus on these two cases (using the shorthands ``Z cut'' and ``XY cut'' in the following), as well as on the charge-SU(2) symmetric line (``SU(2) cut''). Furthermore, we set $U=2$, as the intermediate phases of interest vanish for strong $U$ (see \ref{app:Uvar}).

To solve the model, we mostly use the VUMPS (\textit{variational uniform matrix product states}) framework \cite{Zauner-Stauber_2018}, which works directly in the thermodynamic limit. Our code is equipped to exploit both the spin-SU(2) and the charge-SU(2) of the model, whenever it is appropriate (see \ref{app:VUMPSdetails} for more details). The non-Abelian symmetries are encoded directly into the underlying matrix-product states following the approach in \cite{McCulloch_2007}. In the following, we take the hopping amplitude $t$ as the energy scale, so that all energies are given as dimensionless values in units of $t$; and via setting $\hbar\equiv1$, times are measured in units of $t^{-1}$.

\section{Phase diagram}

\begin{figure}
\centering
\includegraphics[width=\columnwidth]{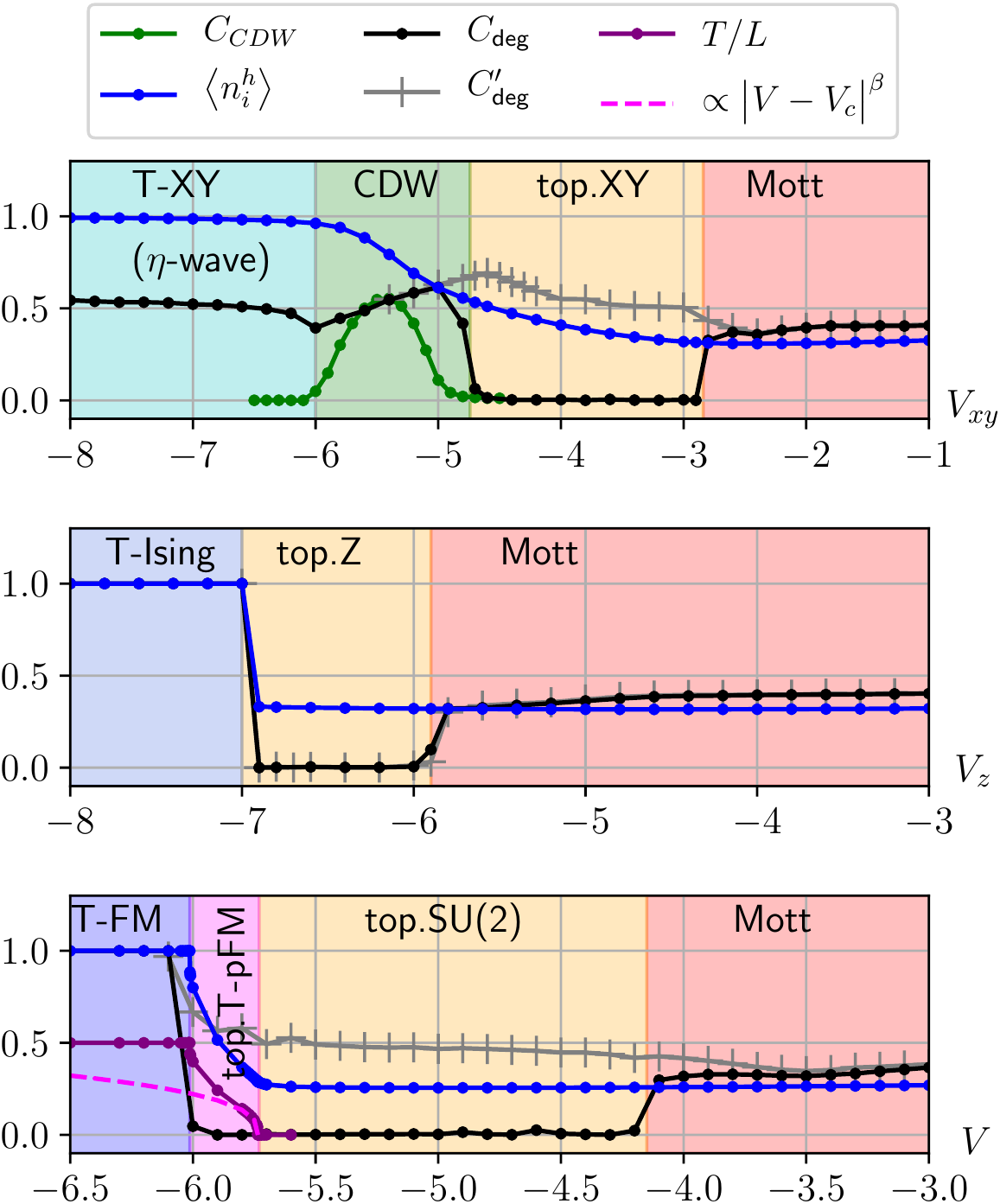}
\caption{\label{fig:PhaseDiagram}
Phase diagram of the model (\ref{eq:model}) for $U=2$ along the cuts $V_{z}=0$ (top), $V_{xy}=0$ (middle) and $V_{xy}=V_z=V$ (bottom), calculated using VUMPS. Displayed is the degeneracy parameter $C_{\textsubscript{deg}}$ of the Haldane phase (\ref{eq:C_Haldane}) ($C'_{\textsubscript{deg}}$ with broken inversion symmetry, see text), the holon density $\avg{n^h_i}=1/L\cdot\partial{\avg{H}}/\partial{U}$ (see text), the total pseudospin density $T/L$ (\ref{eq:T_L}) and the CDW order parameter (\ref{eq:CDW}). The dotted line is a fit of $T/L$ with $\big|V-V_c\big|^{\beta}$ and $\beta\approx0.351$.
}
\end{figure}

It is helpful to consider the extremes of the phase diagram first. For $V_{xy}=V_z=0$, we have the Mott phase, a singlet for both the spin $\v{S}=\sum_i\v{S}_i=0$ (defined as $\v{S}_i = 1/2\sum_{\sigma\sigma'} c^{\dagger}_{i\sigma} \v{\tau}_{\sigma\sigma'} c_{i\sigma'}$, with the Pauli matrices $\v{\tau}$), and the pseudospin $\v{T}=\sum_i\v{T}_i=0$, with a finite charge gap and zero spin gap. If  $V_{xy}<0$ or $V_{z}<0$ is switched on, the Mott phase remains stable in a region that is shaded red in figure \ref{fig:PhaseDiagram}.

For $\big|V_{xy}\big|, \big|V_z\big| \gg t,U$ we are dealing with an XXZ model of pseudospins and can draw from the corresponding knowledge \cite{Mikeska_2004}:
For $\big|V_{xy}\big|>\big|V_z\big|$ the system is in the quasi-long-range-ordered XY-phase (which we call ``T-XY''), with correlations between the $x$- and $y$-components of the pseudospin decaying as $\avg{T^-_0T^+_r} \sim r^{-1/2}$, which translates to long-range pairing correlations $\avg{\cC{0\uparrow}\cC{0\downarrow} \cA{r\downarrow}\cA{r\uparrow}} \sim \lr{-1}^r r^{-1/2}$, interpreted as $\eta$-wave superconductivity \cite{Japaridze_2001}.
For $\big|V_{z}\big|>\big|V_{xy}\big|$, the system is in a
symmetry-broken pseudo-ferromagnetic state $\avg{T^z_i}=\pm1/2$, with the
ground state in the empty ($\avg{n_i}=0$) or fully ($\avg{n_i}=2$)
occupied band (which we call ``T-Ising''). If half filling is forced, one obtains a phase separation between the two configurations \cite{Iemini_2015}, with a domain wall in between.
For $\big|V_{z}\big|=\big|V_{xy}\big|$, the T-XY and the T-Ising phase
mix to form a pseudospin ferromagnet (``T-FM'') with $\avg{\v{T}}=L/2$
which spontaneously breaks the charge-SU(2) symmetry and it becomes
meaningless to distinguish between the two. The ground state still
lies in the empty band, but is now degenerate for all projections
$T^z=\sum_iT^z_i$, i.e. for \textit{all} fillings. Note that spontaneous
symmetry breaking is possible in this 1D system because of the absence of
quantum fluctuations for $\v{T}$, as it holds that $\left[H,\v{T}\right]=0$.

These three superconducting phases are marked by blue hues in figure \ref{fig:PhaseDiagram}. For the Z cut the transition is first-order, for the XY cut it is continuous, with an intermediate charge density wave (CDW) phase (green in figure \ref{fig:PhaseDiagram}) that breaks translational symmetry and can be identified by looking at the order parameter 
\begin{equation}
C_{\textsubscript{CDW}}= \frac{1}{2} \big|\avg{n_i}-\avg{n_{i+1}}\big|.
\label{eq:CDW}
\end{equation}
For the SU(2) cut, we find that the system first passes through a different intervening phase, a partially polarized pseudospin ferromagnet (which we label as ``T-pFM'') with the order parameter given by the pseudospin density $0<T/L<1/2$, where only a range of fillings around half filling is degenerate (see \ref{app:Edeg}). To the best of our knowledge, such a phase has not been reported up to now. We find a second-order transition at $V_{c,1}\approx-5.73$ and $T/L \sim \big|V-V_{c,1}\big|^{\beta}$ with $\beta\approx0.351$, consistent with $\beta=1/3$. An easy way to obtain this T-pFM phase in a matrix-product state framework is by switching off the charge symmetry altogether (we only exploit the SU(2) spin symmetry), allowing for a superposition of different charge states, and by explicitly calculating
\begin{equation}
T/L=\sqrt{\avg{T_i^x}^2+\avg{T_i^y}^2+\avg{T_i^z}^2}.
\label{eq:T_L}
\end{equation}
At the transition to the T-FM phase, the calculation then quickly
converges to the empty or full band. We find a weakly first-order
transition at $V_{c,2}\approx-6.01$ with a small jump in $T/L$ and
$\avg{n^h_i}$.

We come to the main focus of this paper, the topological phases that are marked yellow in figure \ref{fig:PhaseDiagram}. The main evidence for them comes from the two-fold degeneracy of the eigenvalues $s_i$ of the reduced density matrix $\rho_A=\text{Tr}_B\ket{\Psi}\bra{\Psi}$, related to the Schmidt decomposition of the wavefunction into subsystems $A$ and $B$: $\ket{\Psi} = \sum_is_i\ket{\Psi_i^A}\ket{\Psi_i^B}$ (for an infinite MPS, this is always a bipartition at a given bond). In figure \ref{fig:PhaseDiagram}, we plot the staggered sum
\begin{equation}
C_{\textsubscript{deg}} = \sum_i \lr{-1}^i s_i.
\label{eq:C_Haldane}
\end{equation}
This becomes $0$ for even degeneracy, $1$ for a product state, and can
otherwise assume any value in between. 
We find topological phases with $C_{\textsubscript{deg}}=0$ along each of the three cuts and refer to them as ``top.XY'', ``top.Z'' and ``top.SU(2)''.
The phases along the XY and the SU(2) cut are protected
by inversion symmetry only, which can be checked by adding a weak breaking term
$H'=B_{\textsubscript{inv}}\sum_{i\sigma}\lr{-1}^in_{i\sigma}$, $B_{\textsubscript{inv}}=0.01$, that immediately disrupts the full degeneracy (shown as $C'_{\textsubscript{deg}}$ in figure \ref{fig:PhaseDiagram}).
Interestingly, top.Z seems to be protected by more symmetries.
According to our computations, it remains robust even if inversion,
particle-hole and fermion parity symmetry are broken.
Further below, we will also show that the phases are different in terms of correlation functions. Finally, we note that the T-pFM phase also shows $C_{\textsubscript{deg}}=0$, i.e. the system stays topological despite the additional phase transition that leads to superconductivity. 

\section{Edge states} 

To gather further evidence for the topological nature of the phases, we turn to the edge states. Let us once more consider the Haldane chain which hosts entangled $S=1/2$ spins as a guide. They interact, forming a singlet and a triplet \cite{Kennedy_1990}. This means that the correlation between the first and last site is expected to increase compared with the bulk of the chain \cite{Yamamoto_1994, Lang_2015, Iemini_2015b, Ghelli_2020}. On the other hand, if there are no edge states, we expect the correlations between the first site and the rest to simply monotonously decrease with the distance.

\begin{figure}
\includegraphics[width=\columnwidth]{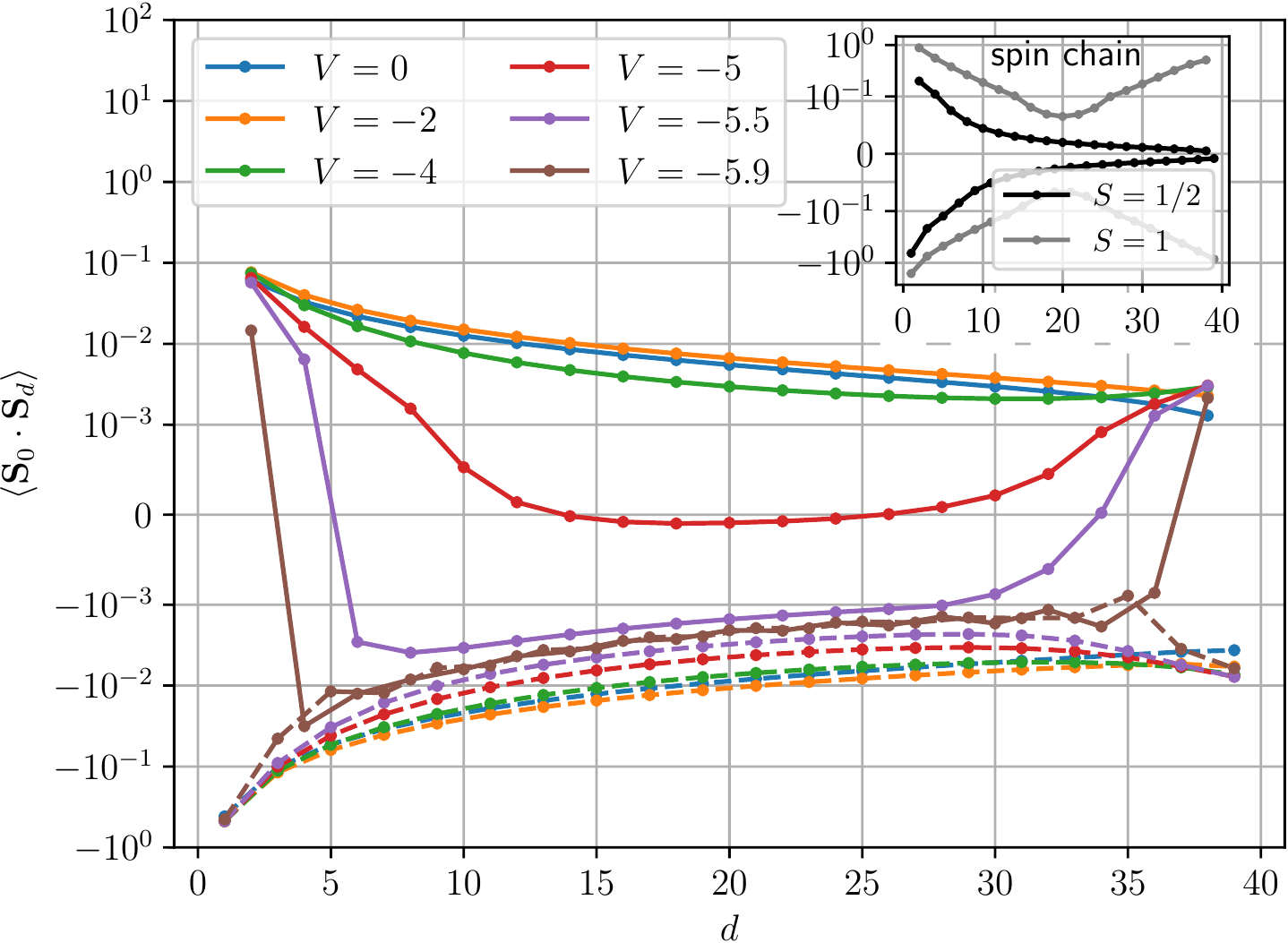}
\caption{\label{fig:Scorr}
Spin-spin correlations in an open chain of length $L=40$ between the
first spin and all the others $\avg{\v{S}_0\cdot\v{S}_d}$ ($d=1\ldots
39$) for $U=2$ and various values of $V=V_{xy}=V_z$. Inset: Comparison with a spin chain $H=\sum_{\left<ij\right>}\v{S}_i\cdot\v{S}_j$ for $S=1/2$ and $S=1$. Expected phases: $V=0,-2,-4$: Mott, $V=-5,-5.5$: top.SU(2), $V=-5.9$: top.T-pFM. Even and odd distances are connected by separate lines as a guide for the eyes.
}
\end{figure}

We test this effect for the SU(2) cut by calculating the spin-spin correlation between the first spin and the rest, displayed in figure \ref{fig:Scorr} for $L=40$ sites. An analogous behavior is found for the other cuts (see \ref{app:edgeStatesXYZ}). The inset compares the nontopological $S=1/2$ spin chain and the topological $S=1$ case. The correlation decays with $d$ in the former case, but has a notable uptick coming close to the opposite edge $d \to L-1$.
The same behavior is found in our fermionic model: While the correlation is clearly monotonically decreasing for $V=0$ and $V=-2$, around $V\approx-4$ a notable uptick starts to develop. We further note that the qualitative behavior shows a crossover from a staggered pattern to mostly antiferromagnetic correlations.

\section{Gaps and excitations} 

\begin{figure}
\includegraphics[width=\columnwidth]{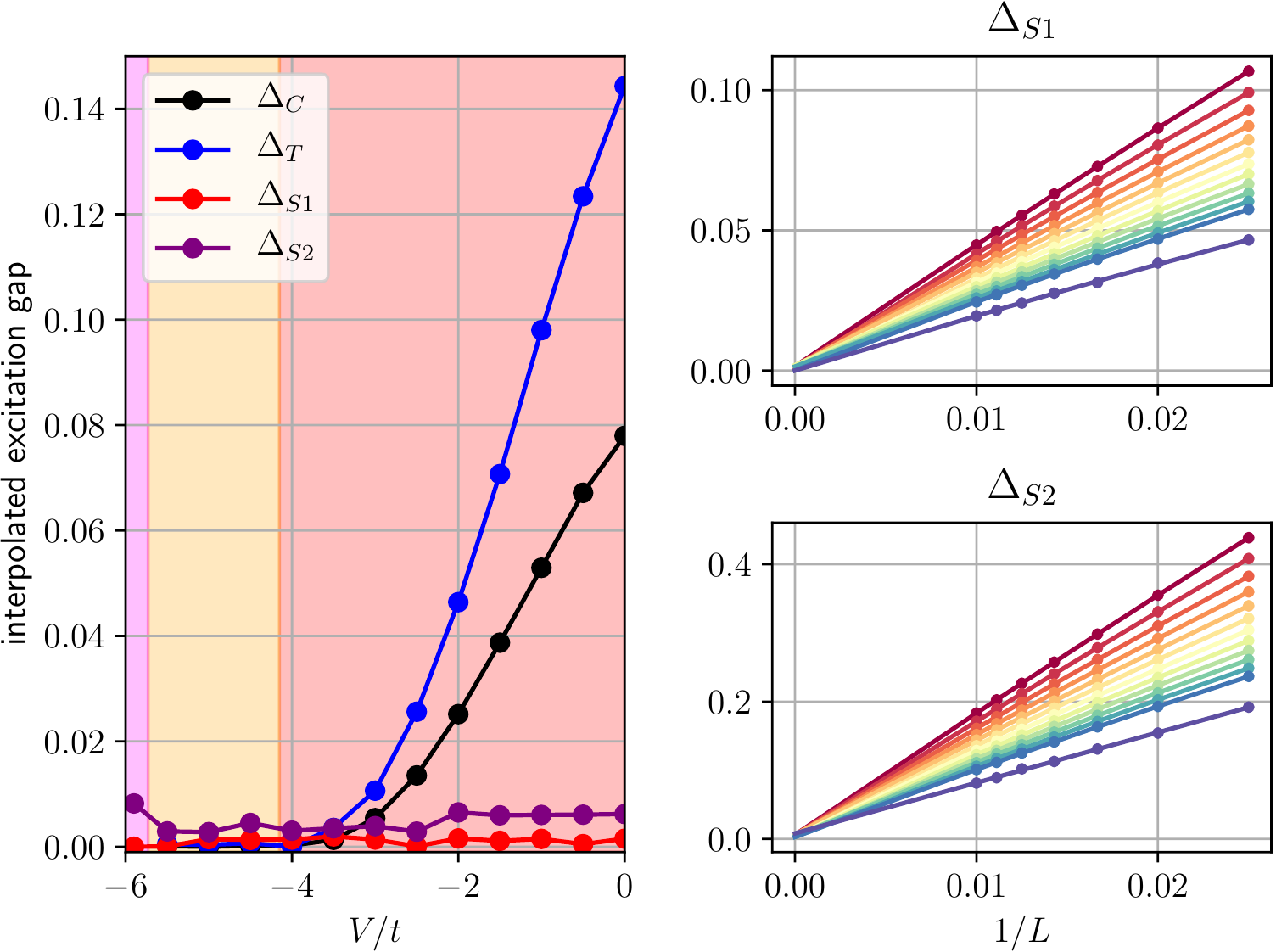}
\caption{\label{fig:gaps}
Gaps of pseudospin (T), charge (C), spin-triplet (S1) and spin-quintet (S2) excitations (see text) for open boundary conditions, calculated with DMRG for chain lengths from $L=40$ to $L=100$ along the charge-SU(2) invariant line $V_{xy}=V_z=V$. The results are interpolated using a second-degree polynomial of $L^{-1}$. The plots on the right show the interpolations of $\Delta_{S1}$ and $\Delta_{S2}$ for (top to bottom) $V=0,-0.5,-1,\ldots,-5,-5.5,-5.9$. The calculations were carried out using SU(2)$\otimes$SU(2) symmetry, except for $V=-5.9$ in the T-pFM phase, where only spin-SU(2) was exploited.
}
\end{figure}

Figure \ref{fig:gaps} shows several excitation gaps along the charge-SU(2) symmetric line of the model: the charge gap $\Delta_{C} = E_0\lr{S=1/2,T=1/2}-E_0\lr{S=0,T=0}$, the pseudospin singlet-triplet gap $\Delta_{T} = E_0\lr{S=0,T=1}-E_0\lr{S=0,T=0}$ (corresponding to the addition or removal of two electrons), the spin singlet-triplet gap $\Delta_{S1} = E_0\lr{S=1,T=T_0}-E_0\lr{S=0,T=T_0}$, and the singlet-quintet gap $\Delta_{S2} = E_0\lr{S=2,T=T_0}-E_0\lr{S=0,T=T_0}$ (corresponding to two spinflips). $T_0$ denotes the pseudospin of the ground state, which is usually $T_0=0$ (i.e. half filling), except for the T-pFM phase, where the pseudospin is partially polarized.

We observe a vanishing of $\Delta_{S1}$, which could be due to the edge states for open boundary conditions as in the Haldane chain \cite{Kennedy_1990}, so that taking $\Delta_{S2}$ into account is also necessary. Surprisingly, we find that $\Delta_{S2}$ vanishes as well, or is at least extremely small. Assuming that the lowest quintet state lies in the continuum of bulk excited states, we have to conclude that the bulk spin gap must vanish. Curiously, the topological phase transition around $V\approx-4.1$ is given by the closing of the charge and the pseudospin gap instead. The closing appears to be exponential, consistent with being of Berezinskii-Kosterlitz-Thouless (BKT) type.

To better understand this behavior, we also look at the dynamics of the bulk system by calculating the spectral function with infinite boundary conditions \cite{Phien_2012}. It is natural to look both at spin excitations given by
\begin{equation}
A_S\lr{k,\omega} = \matrixel{0} { \v{S}_{k\sigma} \delta\lr{\omega + E_{0} - H} \v{S}_{k\sigma} }  {0}
\label{eq:kspecS}
\end{equation}
and at pseudospin excitations given by
\begin{equation}
A_T\lr{k,\omega} = \matrixel{0} { \v{T}_{k\sigma} \delta\lr{\omega + E_{0} - H} \v{T}_{k\sigma} }  {0},
\label{eq:kspecT}
\end{equation}
using the Fourier transform $O_k=1/\sqrt{L} \sum_i \exp\lr{-ikR_i} O_i$ with $O_i=\v{S}_i,\v{T}_i$. In the T-pFM phase, the charge-SU(2) symmetry is reduced to U(1), and we have to look at the individual components, e.g. $O_i= T^z_{i}$.

\begin{figure*}
\includegraphics[scale=0.8]{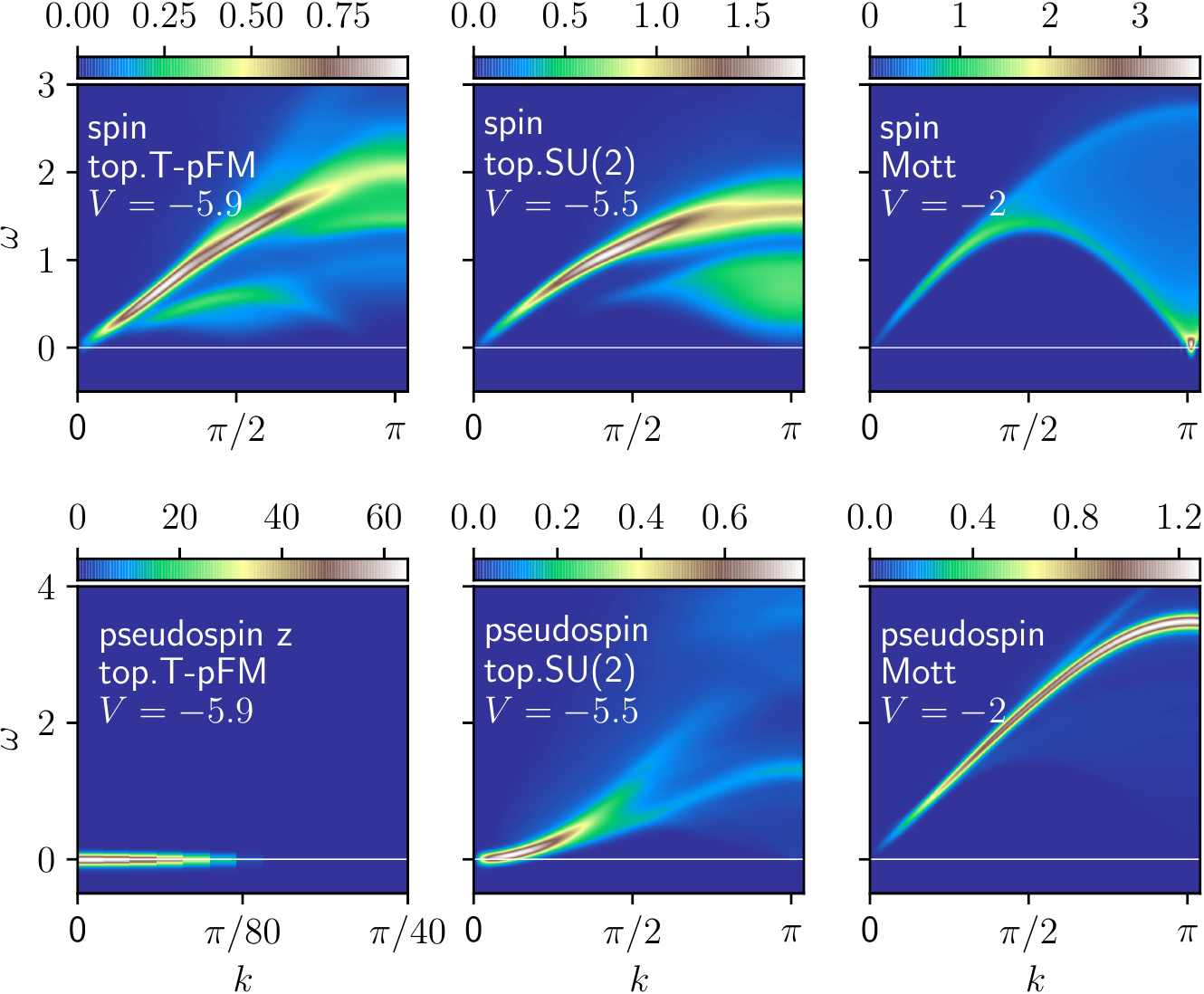}
\caption{\label{fig:spec}
Dynamical spin and pseudospin structure factor at the charge-SU(2) invariant line, for $U=2$, values of $V=V_{xy}=V_z$ and phases as indicated. The pseudospin is approximately half-polarized $T/L\approx0.24$ in the T-pFM phase. The spin structure factor is according to equation (\ref{eq:kspecS}), while the pseudospin structure factor is according to equation ($\ref{eq:kspecT}$), except for the T-pFM phase, where charge-SU(2) is broken and we use only the $z$-component. Additional parameters: infinite boundary conditions with a heterogenous section of length $L=160$, maximal propagation time $t_{\textsubscript{max}}=48$ inverse hoppings before taking the Fourier transform.
}
\end{figure*}

The result is displayed for the SU(2) cut in figure \ref{fig:spec}. One observes that there is in fact a small gap at $k=\pi$ in the top.SU(2) phase and it becomes large at the transition to top.T-pFM. At $k=0$, the spin excitations seem to be gapless, but show a kind of pseudogap behavior, with the spectral weight going to zero for $\omega\to0$. We conclude that these features appear to be enough to protect the topology. The same behavior is found for the other two cuts (see \ref{app:specXYZ}).

Looking at the pseudospin excitations in figure \ref{fig:spec}, one observes that they are slightly gapped in the Mott phase for $V=-2$, while the gap has closed at $V=-5.5$. At $V=-5.9$, the pseudospin is polarized and we obtain an intense (pseudo-)ferromagnetic peak at $k=0$, $\omega=0$. Thus, the topological features in the spin degrees of freedom can coexist with various charge orders in this system.

\begin{figure}
\includegraphics[width=\columnwidth]{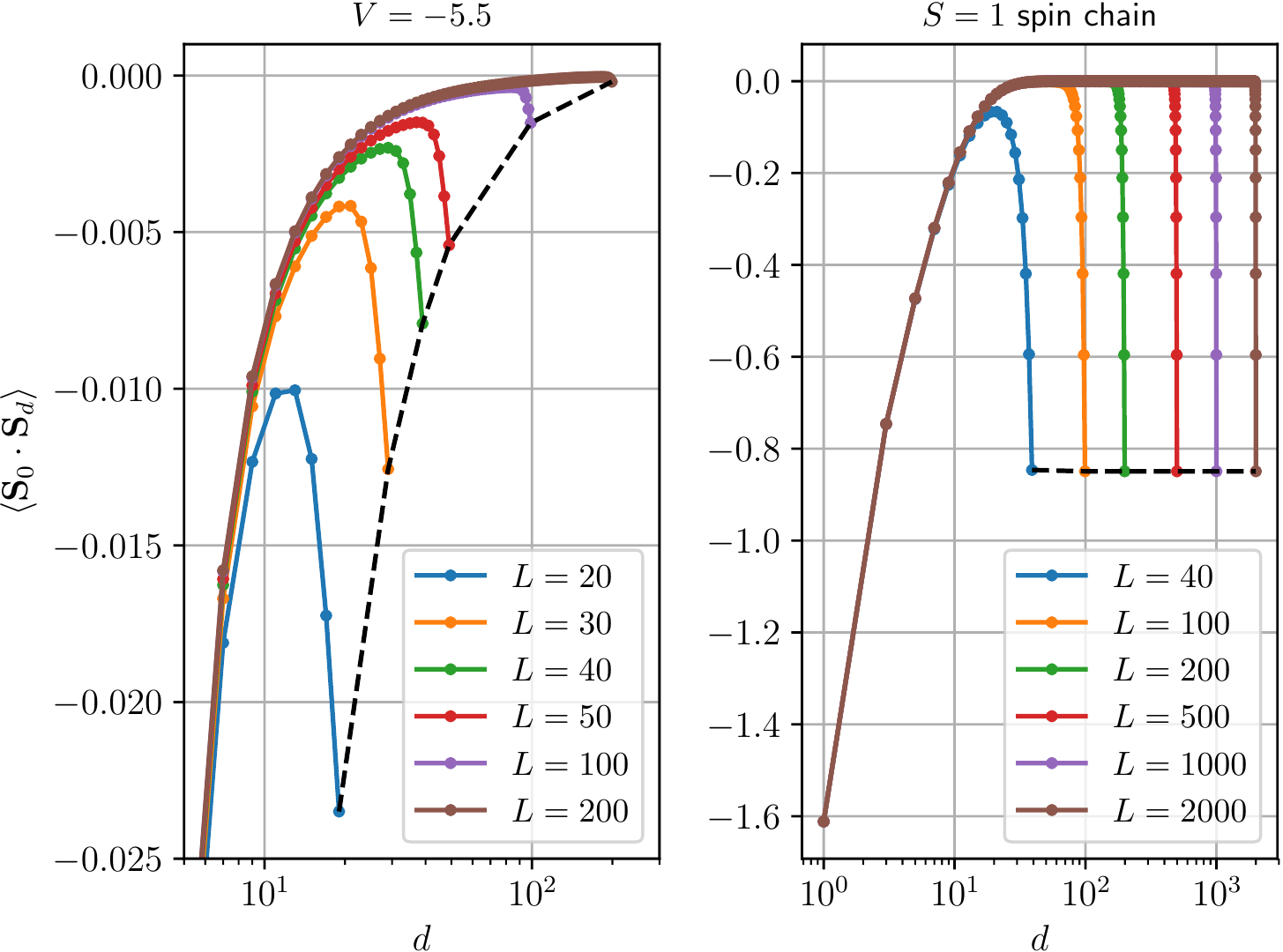}
\caption{\label{fig:Scorr_Ldep}
Correlation function between the first spin and all the rest $\avg{\v{S}_0\cdot\v{S}_d}$ ($d=1\ldots
L-1$) as in figure \ref{fig:Scorr}, now for various chain lengths $L$. Left: Model (\ref{eq:H_SU2}) for $V=-5.5$. Right: $S=1$ spin chain. Note that only odd distances are shown to avoid plot clutter. The black dotted line indicates how strongly the first and last site are correlated as a function of $L$.
}
\end{figure}

While being gapelss does not preclude topological edge states in principle \cite{Keselman_2015, Verresen_2019, Lang_2015, Iemini_2015b}, we may wonder whether they are in any way less robust than in the gapped case (where excitations across the bulk are exponentially suppressed). To investigate this, we return to $\avg{\v{S}_0\cdot\v{S}_d}$ for open boundary conditions and now look at it as a function of the chain length $L$. The result is displayed in figure \ref{fig:Scorr_Ldep} and compared to the $S=1$ spin chain. The revival of this function is quite dramatic in the latter case and the correlation between the first and last site remains constant even for very large system sizes. In our fermionic model it is much more modest and we find that the correlation between the first and the last site decreases approximately as $L^{-1.5}$. The absolute value is also at least an order of magnitude smaller, even when adjusted for the smaller value of the spin. In this sense, we are indeed dealing with weaker and less robust edge states, which is likely a result of the vanishing spin gap.

\section{Correlation functions}

\begin{figure}
\includegraphics[width=\columnwidth]{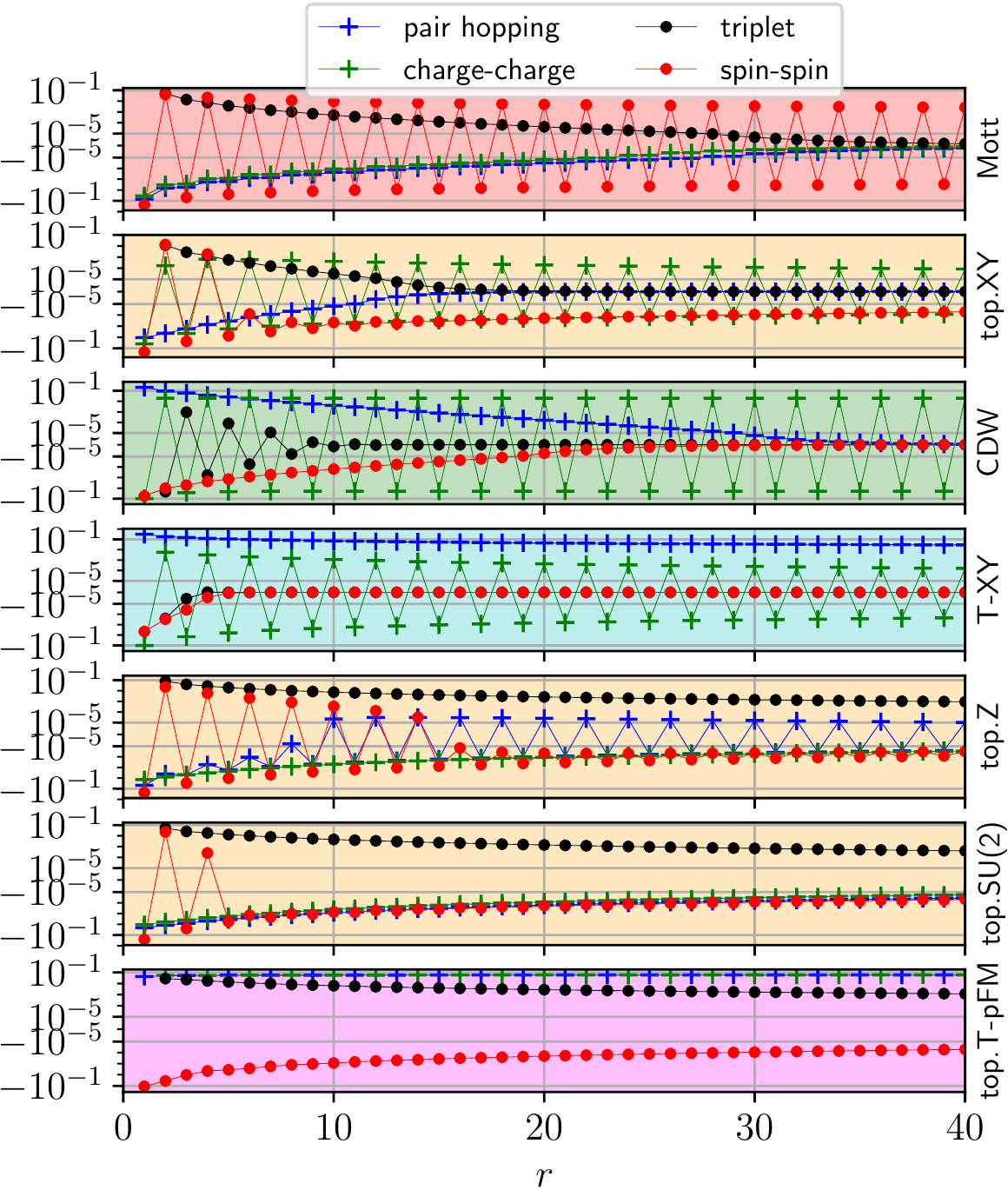}
\caption{\label{fig:corr}
Correlation functions for selected points in the various phases of figure \ref{fig:PhaseDiagram}. 
Mott: $V_{xy}=V_z=0$;
top.XY: $V_{xy}=-3.6$, $V_z=0$;
CDW: $V_{xy}=-5.4$, $V_z=0$;
T-XY: $V_{xy}=-7$, $V_z=0$;
top.Z: $V_{xy}=0$, $V_z=-6.9$;
top.SU(2): $V=-5.5$;
top.T-pFM: $V=-5.9$.
The correlation functions are: pair hopping: $1/2\lr{\avg{T^+_0T^-_r}+\avg{T^-_0T^+_r}}$, charge-charge: $\avg{T^z_0T^z_r}$; triplet-triplet: $\avg{\tau^{\dagger}_0\tau_r}$ with $\tau_r=c_{r\uparrow}c_{r+1,\downarrow}+c_{r\downarrow}c_{r+1,\uparrow}$; spin-spin: $\avg{\v{S}_0\cdot\v{S}_r}$. In the charge-SU(2)-invariant phases, pair-hopping and charge-charge correlations are replaced by $\avg{\v{T}_0\cdot\v{T}_r}$.
}
\end{figure}

In figure \ref{fig:corr} we show correlation functions at selected points within the various phases. Curiously, the topological phases are characterized by short-range AFM correlations up to a certain length and all-negative correlations beyond that. We find that as $\big|V_{xy}\big|$ or $\big|V_{z}\big|$ are increased, the antiferromagnetic range shrinks, and correspondingly the gap at $k=\pi$ of the spin-spin spectral function increases.
However, the phases are different in their charge properties: In the top.XY phase we find a staggered quasi-long-range order in the charge-charge correlations (a precursor of the eventual CDW), but decaying triplet and pair hopping correlations, while the top.Z and top.SU(2) phases show quasi-long-range order in the latter two.
As soon as the spin-spin correlations turn all negative, there is a transition to a true long-range ordered state, which is nontopological CDW in the case of the XY cut, but topological T-pFM in the case of the SU(2) cut.

\section{Conclusions}

We have shown that the Hubbard chain with attractive density-density interaction and/or positive pair-hopping hosts topological phases for the spin degrees of freedom which can coexist with various orders in the charge sector, in particular with long-range $\eta$-wave superconductivity.
This is an unusual instance of topological order arising from interactions and exhibits unexpected properties: The topological properties arise despite the vanishing spin gap (although there is a selective gap at $k=\pi$ and vanishing spectral weight for $k=0$). The ground state is not dimerized. Furthermore, we observe puzzling all-negative spin-spin correlations beyond a certain length scale, meaning that any given spin tends to align itself antiferromagnetically to all the others.

Clearly, more work needs to be done in order to better understand the results. In particular, we could not establish all the protecting symmetries in the $V_{xy}=0$ case. Furthermore, an intuitive understanding of the nature of the edge states and why they appear is desirable. A wealth of different techniques has been recently applied to both the extended Hubbard model and to analyzing topological order, which should prove fruitful to further diagnose this problem. Recently, a framework was suggested to analyze gapless topological phases in terms of their symmetry properties \cite{Verresen_2019} that could also be applied to our system.

So far, a common denominator of interacting topological phases often seems to be either superconducting \cite{Kraus_2013, Lang_2015, Iemini_2015b} or ferromagnetic-type coupling \cite{Lange_2015, Keselman_2015, Agrapidis_2019}, thus adding a topological twist to the old competition of magnetism and superconductivity. Notably, the presence of a gap is much less of a requirement than in the case of free-electron topological insulators. Another route are topological dimerized phases with a larger unit cell \cite{Gonzalez-Cuadra_2019A, Gonzalez-Cuadra_2019B, Koudai_2019, Ghelli_2020} or systems with a larger unit cell by construction \cite{Nourse_2016}. The extended Hubbard model hosts a nontopological dimerized bond-order wave in the repulsive parameter regime \cite{Nakamura_1999, Vojta_1999, Nakamura_2000} and an intriguing question is whether it can be made topological.

All of these findings can help guide the search for further instances of correlation-induced symmetry-protected topological order with novel properties and we hope that our work constitutes a step towards their understanding and classification.

\acknowledgments
We thank Norio Kawakami and Michael Potthoff for helpful
discussions. R.R. thanks the Japan Society for the Promotion of
Science (JSPS) and the Alexander von Humboldt Foundation. Computations
were partially performed at the Yukawa Institute for Theoretical
Physics, Kyoto; partially at the ISSP computation cluster in the
University of Tokyo; and partially at the PhysNet computation cluster at Hamburg University.
R.R. gratefully acknowledges support by JSPS, KAKENHI Grant No. JP18F18750.

\appendix
\setcounter{section}{0}

\section{Away from the charge-SU(2) symmetric line}

\subsection{Edge states}
\label{app:edgeStatesXYZ}

\begin{figure}
\includegraphics[width=\columnwidth]{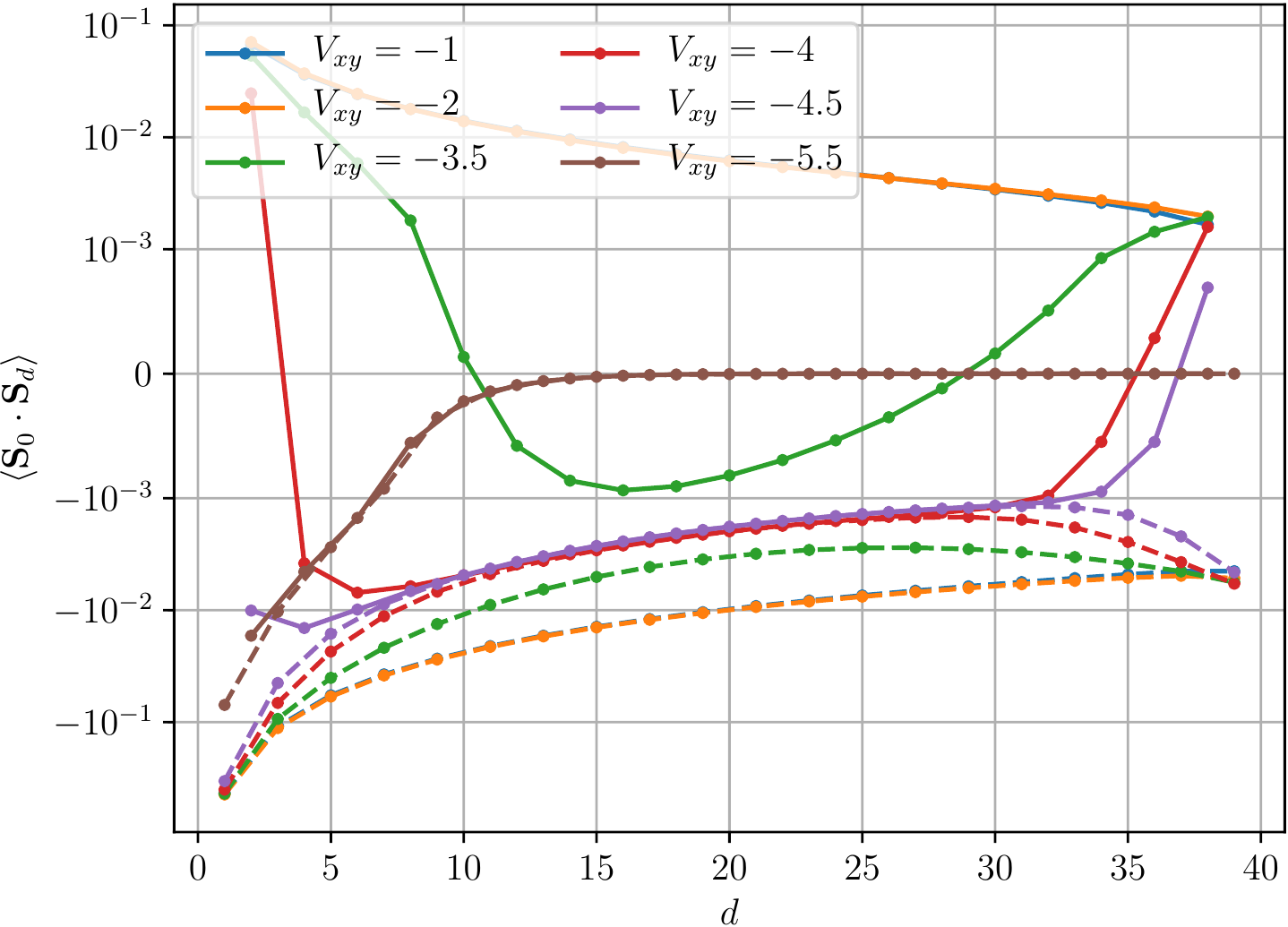}
\caption{\label{fig:ScorrXY}
Spin-spin correlations in an open chain of length $L=40$ between the first spin and all the others $\avg{\v{S}_0\cdot\v{S}_d}$ ($d=1\ldots 39$) for $U=2$, $V_{z}=0$ and various values of $V_{xy}$. Expected phases: $V_{xy}=-1,-2$: Mott, $V_{xy}=-3.5,-4,-4.5$: top.XY, $V_{xy}=-5.5$: CDW. Even and odd distances are connected by separate lines as a guide for the eyes.
}
\end{figure}

Figure \ref{fig:ScorrXY} shows the correlation between the first spin of an open chain and the rest for the XY cut. Just as for the SU(2) cut presented in the main text (figure \ref{fig:spec}), one observes an uptick of the correlation with the last sites, clearly visible for $V_{xy}=-3.5, -4, -4.5$. This is consistent with the position of the topological phase whose limits were obtained from the two-fold degeneracy of the entanglement spectrum of the infinite system (figure \ref{fig:PhaseDiagram}). The correlations decrease monotonously for $V_{xy}=-1,-2$ in the Mott phase and go to zero exponentially for $V_{xy}=-5.5$ in the spin-gapped CDW phase.

\begin{figure}
\includegraphics[width=\columnwidth]{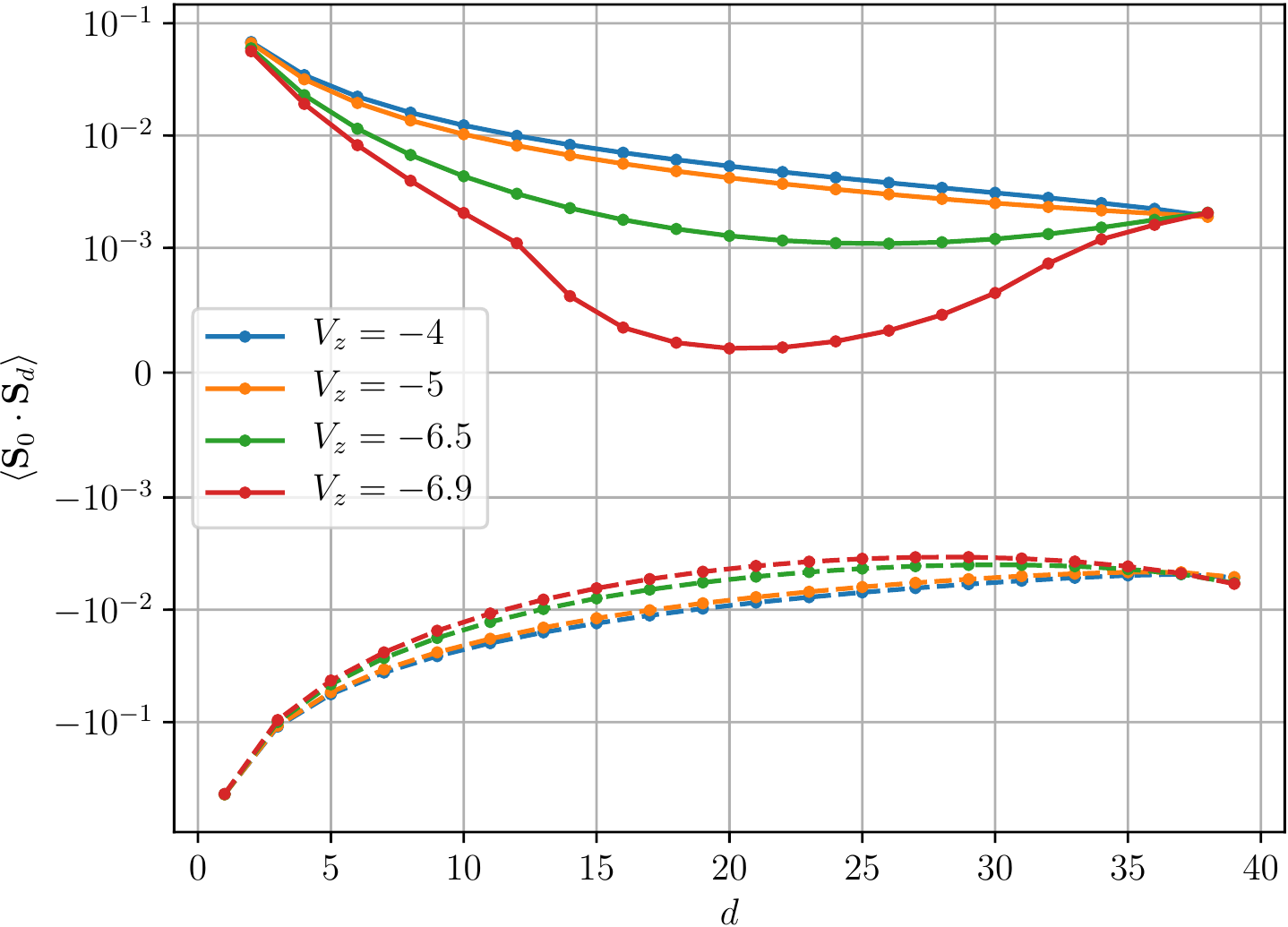}
\caption{\label{fig:ScorrZ}
Same as figure \ref{fig:ScorrXY}, but for $V_{xy}=0$ and various values of $V_z$. Expected phases: $V_z=-4,-5$: Mott, $V_z=-6.5,-6.9$: top.Z. Even and odd distances are connected by separate lines as a guide for the eyes.
}
\end{figure}

Figure \ref{fig:ScorrZ} shows the same for the Z cut, where the uptick is visible for $V_z=-6.5,-6.9$, again consistent with the phase diagram, though the behavior seems somewhat more shallow in this case. One needs to go very deep into the phase (close to the critical $V_z\approx-7$) to see it.

\subsection{Spectral functions}
\label{app:specXYZ}

\begin{figure*}
\centering
\includegraphics{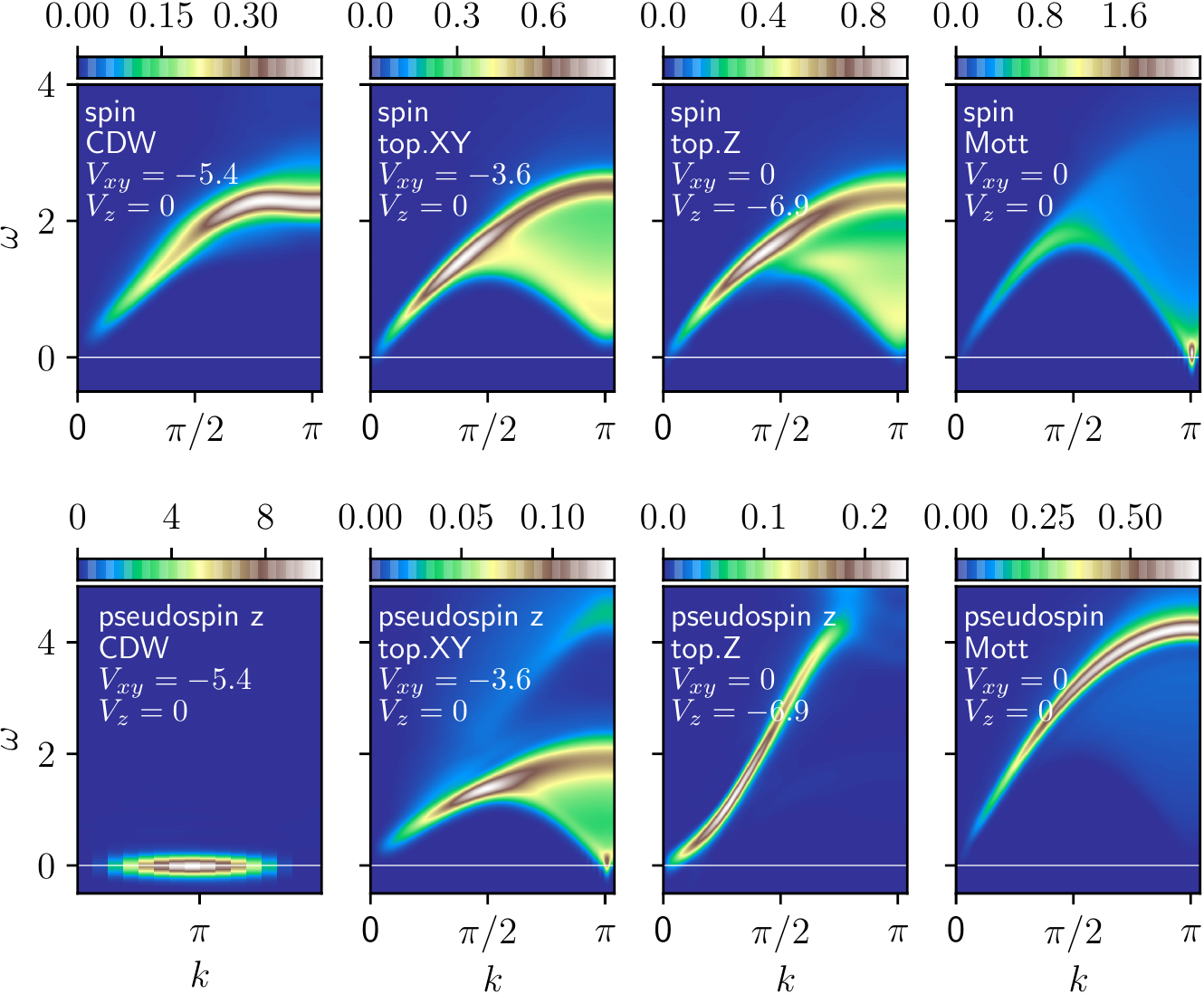}
\caption{\label{fig:specXYZ}
Dynamical spin and pseudospin structure factor (only the z component where indicated) for the parameters and phases as shown. Additional parameters: infinite boundary conditions with a heterogenous section of length $L=160$, maximal propagation time $t_{\textsubscript{max}}=24$ inverse hoppings before taking the Fourier transform.
}
\end{figure*}

Figure \ref{fig:specXYZ} shows the spin and pseudospin spectral function in the topological phases for the XY and for the Z cut, comparing with the Mott case ($V_{xy}=V_z=0$) and the CDW case. We note that the spin spectral function in the topological phases (center two columns) exhibits the same qualitative behaviour as for the SU(2) cut shown in figure \ref{fig:spec}: The strong antiferromagnetic peak at $k=\pi$ dissolves, leaving a small gap at $k=\pi$ and a pseudogap-like suppresion of spectral weight at $k=0$.

The pseudospin excitations are in both cases qualitatively very different: In the top.Z phase, the low-energy excitations are similar to the Mott phase, but gapless, though with vanishing weight for $\omega\to0$. In the top.XY phase, they show a (pseudo-)antiferromagnetic behavior with a strong gapless peak at $k=\pi$, corresponding to quasi-long-range order in the static charge-charge correlation shown in figure \ref{fig:corr}. This is due to $V_{xy}<0$ being equivalent to a repulsive doublon-doublon interaction, favoring configurations with alternating empty and doubly occupied sites. As $\big|V_{xy}\big|$ is increased further ($V_{xy}=-5.4$), it leads to a CDW phase, i.e. a true long-range ordering in the $T^z$ component that shows up as an intense peak at $k=\pi$, $\omega=0$; and eventually to $\eta$-wave superconductivity of doublons in the T-XY phase (not shown).

\section{Details of the VUMPS calculation}
\label{app:VUMPSdetails}

In our VUMPS algorithm implementation we start with a small bond dimension and increase it dynamically once the variation error and the state error have sufficiently converged. The resulting effective bond dimension $\chi$ typically reaches values of  $\chi\sim6.5\cdot 10^3$ when only spin-SU(2) is exploited (in the T-pFM phase), $\chi\sim10\cdot 10^3$ when SU(2)$\otimes$U(1)  is exploited (for $V_{xy}\neq V_z$), and $\chi\sim20-40\cdot 10^3$ when full SU(2)$\otimes$SU(2) is exploited (for $V_{xy}=V_z$).

However, to correctly obtain the degeneracies of the eigenvalue spectrum it seems that a certain symmetry breaking is necessary. This can be checked for the simpler case of the $S=1$ spin chain: When the singular values are resolved by the magnetic quantum number $M$, the first degenerate pair might be found for $M=0$ and $M=1$, the next for $M=-1$ and $M=2$ and so on, where the exact position is random. While this is easily obtainable in the spin chain, we find it is more difficult for our fermionic model, even though all the correlation functions (which are proper observables) converge. We find that singular value degeneracy in the topological phase is quickly reached either without any symmetries at all or only with one U(1) symmetry. Therefore, the degeneracy parameter $C_{\textsubscript{deg}}$ in the main text is calculated for spin-U(1) only, with a bond dimension of around $\chi\sim1.2\cdot10^3$, while we use the maximal symmetries for all other calculations.

\section{Degeneracy close to half filling in the T-pFM phase}
\label{app:Edeg}

Table \ref{tab:energies} shows the ground-state energies for various fillings in the T-pFM phase for $L=40$ and $V=-5.9$, calculated with SU(2)$\otimes$U(1) symmetry, about $100$ half-sweeps, resulting in an energy variance per site $\lr{\avg{H^2}-E^2}/L$ of the order of $10^{-6}$. We see that the energies are near-degenerate, with a difference only in the 6-th digit down to a filling of $n=0.6$. We expect a complete degeneracy in the thermodynamic limit between $n=1$ and $n\approx0.52$.

\begin{table}
\centering
\begin{tabular}{llllll}
$N$ & $n$ & $E_0/L$ & $N$ & $n$ & $E_0/L$\\
\toprule
40 	&	1.0 	&	-0.476414 &	22 	&	0.55 	&	-0.476188\\
38 	&	0.95 	&	-0.476414 &	20 	&	0.5 	&	-0.475594 \\
36 	&	0.9 	&	-0.476414 &	18 	&	0.45 	&	-0.474608\\
34 	&	0.85 	&	-0.476414 &	14 	&	0.35 	&	-0.471431 \\
32 	&	0.8 	&	-0.476414 &	16 	&	0.4 	&	-0.473220 \\
30 	&	0.75 	&	-0.476415 &	12 	&	0.3 	&	-0.469258 \\
28 	&	0.7 	&	-0.476415 &	10 	&	0.25 	&	-0.466741 \\
26 	&	0.65 	&	-0.476415 &	8 	&	0.2 	&	-0.463981 \\
24 	&	0.6 	&	-0.476415 &	6 	&	0.15 	&	-0.460746 \\
	&	&	&	4 	&	0.1 	&	-0.457503 \\
	&	&	&	2 	&	0.05 	&	-0.448273 \\
	&	&	&	0 	&	0.0 	&	-0.438125 \\
\end{tabular}
\caption{\label{tab:energies}
Ground-state energies in all the particle number sectors for $L=40$, $U=2$, $V=-5.9$, corresponding to a pseudospin polarization of $T/L\approx0.24$ in the thermodynamic limit.
}
\end{table}

\section{Varying $U$}
\label{app:Uvar}

\begin{figure}[!th]
\centering
\includegraphics[width=\columnwidth]{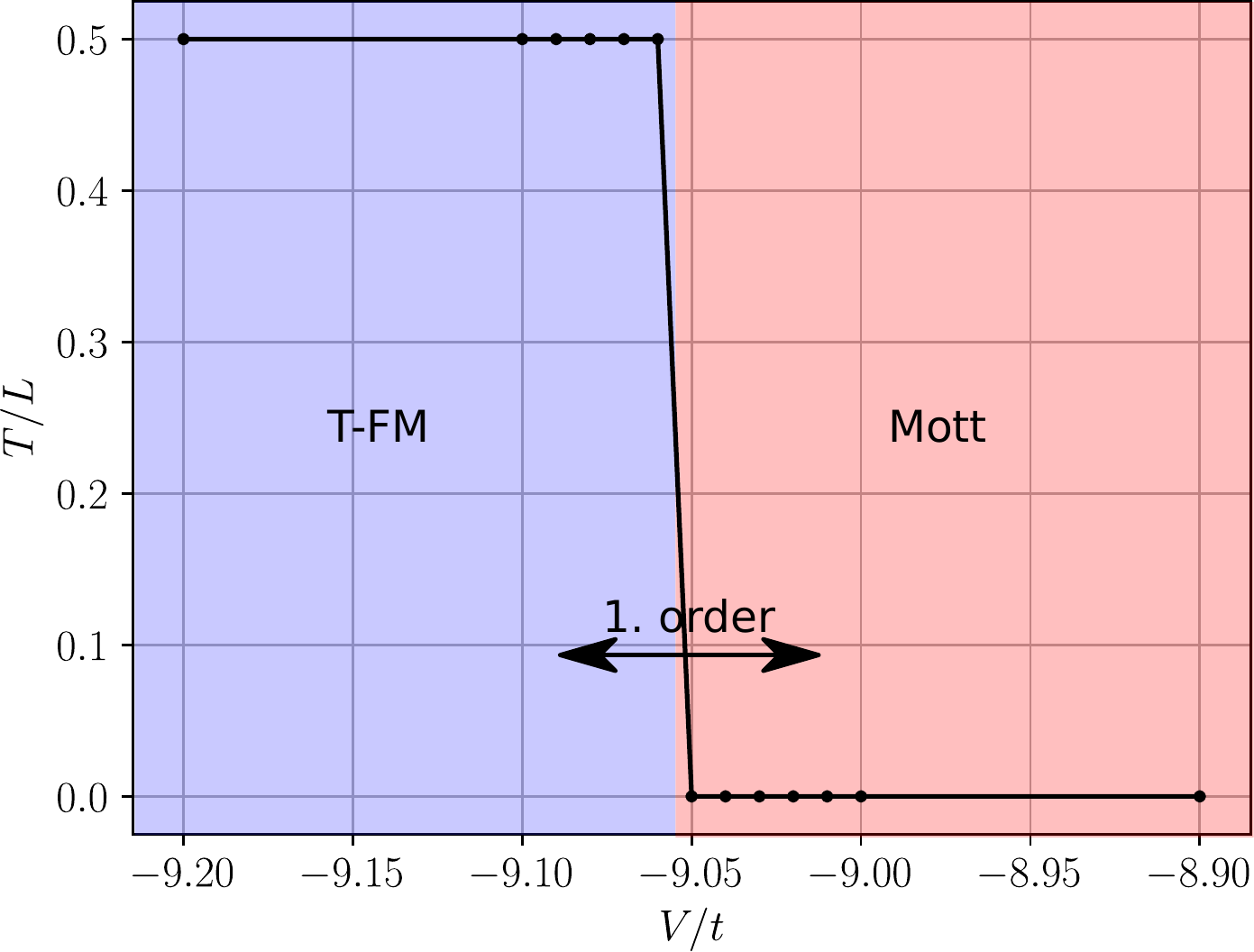}
\caption{\label{fig:U=4}
Phase diagram along the SU(2) cut $V=V_{xy}=V_z$ for $U=4$, taking the pseudospin density $T/L$ as order parameter.
}
\end{figure}

Figure \ref{fig:U=4} shows the phase diagram along the SU(2) cut for $U=4$. We find that the intervening phases disappear and there is just a first-order phase transition to the T-FM phase at $V_c\approx9.055$.

To understand this it is helpful to consider vanishing hopping $t=0$ in our model (\ref{eq:model}). In this case, we are just left with the $U$-term and the $V$-term, which commute. The former favors a state with uniform single occupancy and an energy $E=0$, while the latter favors the empty band (ferromagnetically aligned pseudospins) with an energy of $E=U/2+V/4$. The two lines cross at $V=-2U$ where a first-order transition takes place due to a level crossing.
Thus, the presence of the interesting intervening phases is an effect of non-negligible hopping, i.e. they appear for $U\sim t$ and the corresponding transition lines must end at a critical endpoint $U_c$ beyond which the transition is first order. We estimate $U_c\approx2.25$.

\FloatBarrier

%

\end{document}